\documentclass[twoside,twocolumn,english,prl,aps,superscriptaddress,showpacs]{revtex4}
\usepackage[T1]{fontenc}
\usepackage[latin9]{inputenc}
\usepackage{color}
\usepackage{babel}
\usepackage{textcomp}
\usepackage{amsmath}
\usepackage{mathptmx}
\usepackage{amssymb,amsfonts,mathrsfs}
\usepackage{amsthm}
\usepackage{graphicx}
\usepackage[unicode=true,pdfusetitle,
bookmarks=true,bookmarksnumbered=false,bookmarksopen=false,
breaklinks=false,pdfborder={0 0 0},backref=false,colorlinks=true]
{hyperref}

\usepackage{breakurl}
\usepackage{epstopdf}
\usepackage{hyperref}

\makeatother
\theoremstyle{plain}

\hypersetup{
    colorlinks=ture,
    linkcolor=red
}
\makeatother
\begin{document}
	
\preprint{This line only printed with preprint option}

\title{Ring Structure in the Complex Plane: A Fingerprint of non-Hermitian Mobility Edge}

\author{Shan-Zhong Li}
\affiliation {Key Laboratory of Atomic and Subatomic Structure and Quantum Control (Ministry of Education), Guangdong Basic Research Center of Excellence for 
Structure and Fundamental Interactions of Matter, School of Physics, South China Normal University, Guangzhou 510006, China}

\affiliation {Guangdong Provincial Key Laboratory of Quantum Engineering and Quantum Materials, Guangdong-Hong Kong Joint Laboratory of Quantum Matter, Frontier Research Institute for Physics, South China Normal University, Guangzhou 510006, China}

\author{Zhi Li}
\email{lizphys@m.scnu.edu.cn}

\affiliation {Key Laboratory of Atomic and Subatomic Structure and Quantum Control (Ministry of Education), Guangdong Basic Research Center of Excellence for 
Structure and Fundamental Interactions of Matter, School of Physics, South China Normal University, Guangzhou 510006, China}

\affiliation {Guangdong Provincial Key Laboratory of Quantum Engineering and Quantum Materials, Guangdong-Hong Kong Joint Laboratory of Quantum Matter, Frontier Research Institute for Physics, South China Normal University, Guangzhou 510006, China}

\date{\today}

\begin{abstract}
By Avila's global theory, we analytically reveal that the non-Hermitian mobility edge will take on a ring structure in the complex plane, which we name as ``mobility ring''. The universality of mobility ring has been checked and supported by the Hermitian limit, $PT$-symmetry protection and without $PT$-symmetry cases. Further, we study the evolution of mobility ring versus quasiperiodic strength, and find that in the non-Hermitian system, there will appear multiple mobility ring structures. With cross-reference to the multiple mobility edges in Hermitian case, we give the expression of the maximum number of mobility rings. Finally, by comparing the results of Avila's global theorem and self-duality method, we show that self-duality relation has its own limitations in calculating the critical point in non-Hermitian systems. As we know, the general non-Hermitian system has a complex spectrum, which determines that the non-Hermitian mobility edge can but exhibit a ring structure in the complex plane.
\end{abstract}

\maketitle

\textcolor{blue}{\emph{Introduction.}}---In 1958, P.W. Anderson published a groundbreaking finding that the wave function of electrons can become exponentially localized under the influence of disorder, which was soon well-known as Anderson localization~\cite{PWAnderson1958}. In low-dimensional (1D and 2D) disordered systems, the presence of disorder, no matter how small it is, will cause all eigenstates in the system to be localized. That is to say, the metal-insulator phase transition induced by change of disorder strength doesn't exist in the 1D or 2D system~\cite{EAbrahams1979,PALee1985,FEvers2008}. However, in the 3D case, the disorder allows the extended state and the localized state to coexist, separated by the mobility edge (ME) of the critical energy~\cite{NFMott1967}. This means that by controlling the strength of disorder or Fermi energy level, metal-insulator phase transitions can be achieved in 3D disordered systems.

Unlike random disordered systems, a quasiperiodic system, as a system between order and disorder, can give rise to ME through controllable metal-insulator phase transition at low dimensions~\cite{YLahini2009,YEKraus2012,MVerbin2013,MVerbin2015,PWang2020,SAGredeskul1989,DNChristodoulides2003,TPertsch2004,GRoati2008,GModugno2010,MLohse2016,SNakajima2016,HPLuschen2018,FAAnK2021,DTanese2014,PRoushan2017,FAAn2018,VGoblot2020,JGao2023}. The 1D Aubry-Andr\'e-Harper (AAH) model we study in this paper is among the most famous quasiperiodic systems~\cite{PGHarper1955,SAubry1980}. As is known, the standard AAH model has a precise critical point of metal-insulator phase transition which does not depend on eigenenergy, therefore no ME can be expected in such systems~\cite{SAubry1980}. However, recent studies have shown that energy-dependent MEs can be induced in 1D AAH models by introducing short-range dimered hopping~\cite{SRoy2021}, long-range hopping~\cite{JBiddle2010,XDeng2019,NRoy2021}, or manipulating the structure of quasiperiodic potentials~\cite{SGaneshan2015, YWang2020, APadhan2022,SDasSarma1988,SDasSarma1990,TLiu2017}.

On the other hand, non-Hermitian Hamiltonians have garnered considerable attention for their ability to effectively describe open or non-conservative systems. Subsequently, novel non-Hermitian phenomena were discovered one after another~\cite{SLonghi2019a,LZhou2024,SZLi2024,YXiong2018,ZGong2018,SYao2018,KKawabata2019,CHLee2019,KYokomizo2019,LXiao2020, KZhang2020, NOkuma2020, DSBorgnia2020, ZYang2020, CXGuo2021, LLi2020, EJBergholtz2021,KKawabata2023,HJiang2019,VVKonotop2016,RElGanniny2018,YAshida2020,CMBender1998,AGuo2009,ARegensburger2012,SWeimann2017,XNi2018,MKremer2019,SXia2021,YLi2022,PPeng2016,JLi2019,ZRen2022,LXiao2017,HZLi2023,XJYu2023,ZXGuo2022,LZhou2023a,KLi2023,RHamazaki2019,XZhang2022,DWZhang2020a,DWZhang2020b,LJLang2021,LJZhai2020}. In general, a non-Hermitian system features complex eigenvalues where the imaginary part corresponds to the eigenstate's lifetime. The difference in the eigenvalues gives non-Hermitian systems a unique ``point gap'' different from Hermitian systems, allowing for a non-zero topological winding number even in a single-band model~\cite{ZGong2018,NOkuma2020}.

However, not all non-Hermitian systems have complex eigenvalues. Recent research shows that if the Hamiltonian satisfies the property of $\eta$-psedo-Hermiticity (including $PT$- symmetry as a special case), the eigenvalues of the system can be guaranteed to be pure real numbers~\cite{CMBender1998,CMBender2007,REIGanainy2018,YAshida2020}. So far, most studies of non-Hermitian MEs focus on systems with $PT$-symmetry, and their results reveal that $PT$-symmetry breaking shares the same boundary as MEs~\cite{SLonghi2021,SLonghi2019,YLiu2021,YLiu2020,TLiu2020,TLiu2022,XXia2022}. This means that the extended state is under the protection of $PT$-symmetry with the corresponding eigenvalue being a real number, while the localized state is in the $PT$-symmetry breaking phase, corresponding to complex eigenvalues and non-trivial topological point gaps~\cite{SLonghi2019}. This one-to-one correspondence has led to the fact that most previous studies on non-Hermitian MEs were based on real eigenvalues. However, in most cases the eigenvalues of non-Hermitian systems are complex numbers rather than pure real numbers, which naturally brings the question: how does a non-Hermitian ME act as a boundary in the complex plane to separate the extended state from the localized state?

\begin{figure}[tbp]
	\centering	
    \includegraphics[width=7cm]{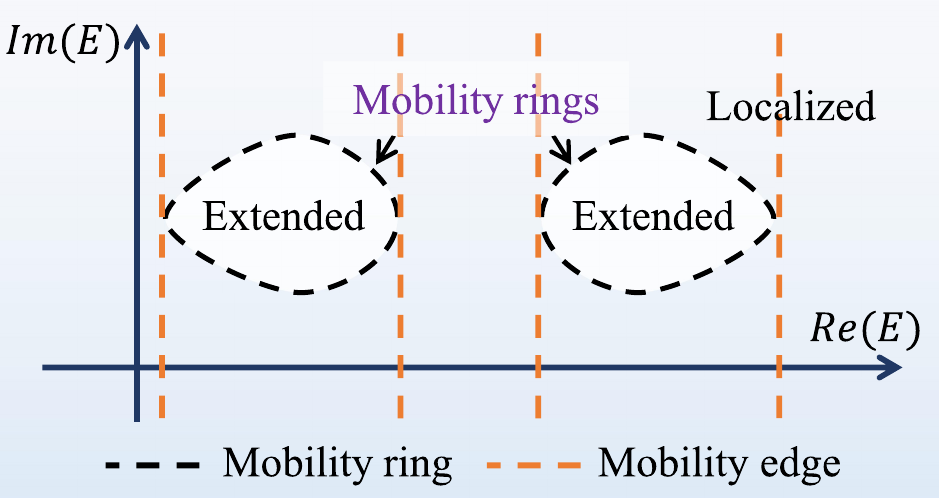}
	\caption{A schematic of MR in the complex plane. The extended (localized) state is inside (outside) MR. ME is essentially a projection of MR on the real axis.}\label{F1}
\end{figure}

In this Letter, we prove analytically that the non-Hermitian ME always presents ring structure in the complex plane, which we name ``mobility ring (MR)''. In order to verify the universality of MRs, we study three precisely solvable cases. By leveraging Avila's global theory, which rigorously characterizes the localization properties of eigenstates through the Lyapunov exponent (LE), we obtain the analytical expression of the corresponding ME. The results show that since LE depends on both the real and imaginary parts of the eigenvalues, MEs form a ring structure in the complex plane, which is MR. MR acts as a clear boundary separating the extended and localized states (see Fig.~\ref{F1}). The emergence of MRs means that there is an infinite number of self-duality points in the complex plane, suggesting that the self-duality relation is no longer suitable for the study of mobility edges in the non-Hermitian case.

%%%%%%
%%%%%%
%%%%%%
%%%%%%

\textcolor{blue}{\emph{Model.}}---We consider a non-Hermitian mosaic quasiperiodic model. The corresponding Hamiltonian reads
%%%
%%%
\begin{equation}\label{Hami}
H=\sum_{j=1}^{L-1}J(c_{j}^{\dagger}c_{j+1}+\mathrm{H.c.})+\sum_{j=1}^{L}V_{j}c_{j}^{\dagger}c_{j}, 
\end{equation}
%%%
%%%
where
%%%
%%%
\begin{equation}\label{mosaic}
V_{j}=
\begin{aligned}
\left\{\begin{matrix}
2\lambda \cos(2\pi\alpha j+\theta+ih),  &j  = n\kappa, \\
\delta, ~&\mathrm{else}.
\end{matrix}\right.
\end{aligned}
\end{equation}
%%%
%%%
$c_{j}^{\dagger}$ ($c_{j}$) creates (annihilates) a fermion on site $j$, and $\mathrm{H.c.}$ stands for the Hermitian conjugate. $J$ represents the nearest neighbor hopping strength. The potential is divided into two parts. The first part is a complex quasiperiodic potential that exists at every $\kappa$ site and is modulated by the following parameters: $\lambda$ denotes the quasiperiodic intensity, $\alpha$ is the quasiperiodic parameter, and $\theta+ih$ denotes a complex phase factor. The remaining lattices have a complex constant potential $\delta$. Since the quasiperiodic potential periodically appears with an interval of $\kappa$, one can consider it as a quasicell of $\kappa$ sublattices, with the system featuring $N = L/\kappa$ quasicells and $n=1,2,\dots,N$ as the quasicell index, where $L$ is the system size. The non-Hermitian is controlled by $h$ and $\delta$. When $\kappa = 1$, $\delta = 0$, and $h = 0$, the model returns to the AAH model, and the self-duality relation indicates a localization phase transition at $\lambda = 1$. When $\kappa>1$ and $\delta=0$, the Hamiltonian can describe both Hermitian ($h=0$) and non-Hermitian ($h>0$) mosaic models~\cite{YWang2020,YLiu2021}. It is worth noting that when $\delta$ is real, the system exhibits $PT$-symmetry for $\theta=0$ due to $V_{j} = V_{-j}^{\ast}$. In numerical calculations, we adopt periodic boundary conditions and choose an irrational number $\alpha=F_{m-1}/F_{m}$, where $F_{m}$ is the $m$th Fibonacci number. Unless otherwise specified, we set $\lambda>0$, $h\ge 0$, $L=F_{m}$, and $J=1$ as the unit energy.

%%%%%%
%%%%%%
 \begin{figure*}[thbp]
	\centering	
    \includegraphics[width=14cm]{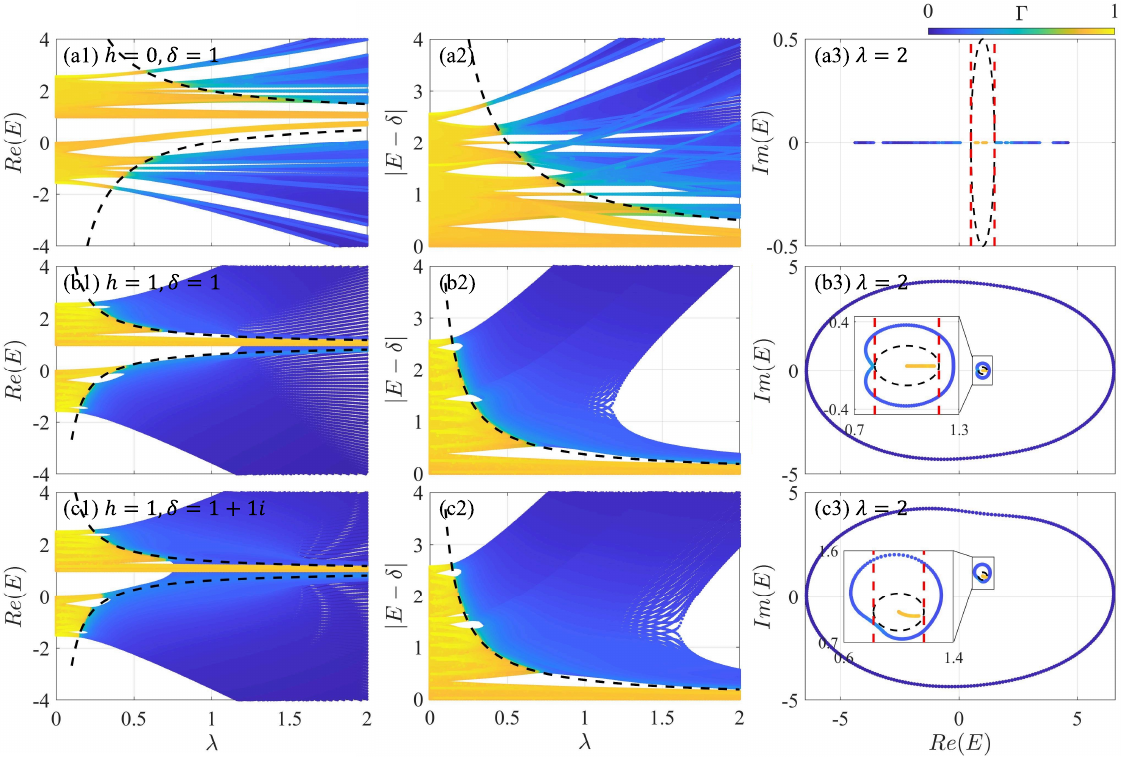}
	\caption{(a1)-(c1) The fractal dimension $\Gamma$ versus $Re(E)$ and $\lambda$, where the dashed lines represent the critical energy $E_{c}=\delta_R\pm\frac{1}{\lambda e^{h}}$. (a2)-(c2) $\Gamma$ versus $|a_2|$ and $\lambda$, where the dashed line is $|a_{2}|=\frac{1}{\lambda e^{h}}$. (a3)-(c3) $\Gamma$ of the eigenstates in complex plane, where the black dashed rings represent the MRs given by Eq.~\eqref{kappa22} and red dashed lines are MEs obtain from $E_{c}=\delta\pm\frac{1}{\lambda e^{h}}$. In numerical computation, we set $h=0$, $\delta=1$ for (a1)-(a3), $h=1$, $\delta=1$ for (b1)-(b3) and $h=1$, $\delta=1+i$ for (c1)-(c3). Throughout, $L=610$ and $\kappa=2$.}\label{F2}
\end{figure*}
%%%%%%
%%%%%%
\textcolor{blue}{\emph{The exact MR.}}---LE is a key observable that reflects the localization properties and MEs, which is defined as
%%%
%%%
\begin{equation}\label{LE}
\gamma_{\epsilon }(E)=\lim_{N\rightarrow \infty }\frac{1}{2\pi N}\int  \ln \left \| T_{N}(\theta )  \right \|d\theta ,
\end{equation}
%%%
%%%
where $T_{N}=\prod_{n=1}^{N} T_{n}=T_{N}T_{N-1}\cdots T_{2}T_{1}$, $T_n$ is the transfer matrix of a quasicell, $N$ is the number of quasicells, and $\left \| \cdot  \right \|$ denotes a matrix norm. In the thermodynamic limit $N\rightarrow\infty$, the extended state has $\gamma = 0$, while the localized state has $\gamma > 0$. The transfer matrix of the model~\eqref{Hami} can be written as
%%%
%%%
\begin{equation}
T_{n}=\begin{pmatrix}
E-V_{j}  & -1  \\
 1 &0
\end{pmatrix}
\begin{pmatrix}
E-\delta  & -1  \\
 1 &0
\end{pmatrix}^{\kappa-1},
\end{equation}
%%%
%%%
where
%%%
%%%
\begin{equation}
\begin{pmatrix}
E-\delta  & -1  \\
 1 &0
\end{pmatrix}^{\kappa-1}=
\begin{pmatrix}
a_{\kappa}  & -a_{\kappa-1}  \\
a_{\kappa-1}  & -a_{\kappa-2}
\end{pmatrix},
\end{equation}
%%%
%%%
and the $a_{\kappa}$ is defined as
%%%
%%%
\begin{equation}
a_{\kappa}=\frac{1}{\sqrt{E'^2-4}}\left[\left(\frac{E'+\sqrt{E'^2-4}}{2} \right)^{\kappa}-\left(\frac{E'-\sqrt{E'^2-4}}{2} \right)^{\kappa} \right]
\end{equation}
%%%
%%%
with $E'=E-\delta$. Further, we employ Avila's global theory of one-frequency analytical $SL(2,\mathbb{C} )$ cocycle~\cite{AAvila2015}. Let $h \rightarrow +\infty$, then direct computation yields
%%%
%%%
\begin{equation}
T_{n,\epsilon\rightarrow +\infty }=\frac{e^{-i(2\pi\alpha j+\theta)}e^{h}}{2}\begin{pmatrix}
-2\lambda a_{\kappa}   & 2\lambda a_{\kappa-1} \\
0 &0
\end{pmatrix}+o(1),
\end{equation}
Thus, we have
%%%
%%%
\begin{equation}
\kappa\gamma_{h\rightarrow +\infty}(E)=
\ln|\lambda a_{\kappa}|+h+o(1).
\end{equation}
%%%
%%%
According to Avila's global theory, as a function of $h$, $\gamma(E)$ is a convex piecewise linear function with integer slopes. For large enough $h$, the slope is 0 or 1. The discontinuity of the slope occurs when $E$ belongs to the spectrum of Hamiltonian $H$ except for $\gamma_{h}(E) = 0$, which represents the extended states. This implies that LE for the system can only be expressed as $\kappa\gamma(E)=\max\left\{\ln|\lambda a_{\kappa}|+h,~0 \right\}$~\cite{YLiu2021}. Since LE is an even function for $h$, we further have
%%%
%%%
\begin{equation}\label{gamma}
    \kappa\gamma(E)=\max\left\{\ln|\lambda a_{\kappa}|+|h|,~0 \right\}
\end{equation}
%%%
%%%
for a localized eigenstate, the localization length is $\Lambda=1/\gamma$, while for an extended eigenstate, $\gamma=0$ corresponds to the localization length $\Lambda \rightarrow\infty$. Thus the MEs are determined by
 %%%
 %%%
\begin{equation}\label{kappa2}
\lambda e^{h}\left| a_{\kappa}\right|=1.
\end{equation}
%%%
%%%
As we cannot guarantee real eigenvalues in non-Hermitian systems, we let $E = E_{R} + iE_{I}$ to further study MR in the complex plane, where $E_{R}$ and $E_{I}$ are the real part and the imaginary part of eigenvalues. When $\kappa=2$, the exact MR is $\lambda e^{h}|a_2|=1$, where $a_2=E_{c}-\delta$. Then, one can obtain the general expression of MEs in the complex plane
%%%
%%%
\begin{equation}\label{kappa22}
x^2+y^2=\frac{1}{(\lambda e^{h})^2},
\end{equation}
%%%
%%%
where $x=E_{R}-\delta_R$ and $y=E_{I}-\delta_I$. It is a circle centered at $(\delta_{R},\delta_{I})$ with radius $1/(\lambda e^{h})$ in $Re(E)-Im(E)$ plane, where $\delta_{R}$ and $\delta_{I}$ serve as the real and imaginary parts of the $\delta$. The eigenstates within the circle possess $\gamma=0$, signifying they are extended states, while the eigenstates outside the circle exhibit $\gamma>0$, indicating they are localized states.

For the case of $\kappa = 3$, ME is given by $\lambda e^{h}|a_3|=1$, where $a_3=(E_{c}-\delta)^2-1$. The corresponding analytical expression of MR in the complex plane reads
%%%
%%%
\begin{equation}\label{kappa3}
(x^2-1)^2+(y^2+1)^2+2x^2y^2-\frac{1}{\lambda^2  e^{2h}}-1=0.
\end{equation}
%%%
%%%
Using the same method, one can also derive analytical expressions for MEs under other values of $\kappa$.

\textcolor{blue}{\emph{Universality of MR}}---To prove the universality of MR, the following three cases will be discussed respectively, i.e., the Hermitian limit ($h=0$, $\delta=1$), the $PT$-symmetric protection ($h>0$, $\delta=1$) and without $PT$-symmetry ($h=1$, $\delta=1+i$) cases. 

Numerically, the localization properties can be reflected by the fractal dimension of eigenstates, which is defined as
%%%
%%%
\begin{equation}
\Gamma(\beta)=-\frac{\ln\xi(\beta)}{\ln L},
\end{equation}
%%%
%%%
where $\xi(\beta)=\sum_{j=1}^{L}[|\psi_{j}(\beta)|^4/|\psi_{j}(\beta)|^2]$ denotes the inverse participation ratio and $\psi_{j}(\beta)$ is the probability amplitude of the $\beta$-th eigenstate at the $j$-th site. $\Gamma\rightarrow 0~(1)$ corresponds to localized (extended) states. 
The fractal dimensions $\Gamma$ corresponding to the Hermitian limit, $PT$-symmetry protection and without $PT$-symmetry cases are plotted in the top, middle and bottom rows of Fig.~\ref{F2}, respectively. The dashed lines and rings correspond to MEs and MRs, respectively.

First, when the imaginary part of eigenvalue is factored out, one can obtain the analytical expression of the critical energy, i.e., $E_{c}=\delta_R\pm\frac{1}{\lambda e^{h}}$. Since the eigenvalues of the Hermitian system are pure real numbers, the analytical results perfectly match the numerical results in Fig.~\ref{F2}(a1) . However, for the non-Hermitian case, whose eigenvalues are complex numbers, ME obtained by only calculating the real part can not see a good match between the analytical and numerical results. In concrete terms, though ME can indicate the extended domain, it fails to delineate clearly the localized region. In other words, ME can ensure that all extended states are within the range $E\in \left(\delta_R-\frac{1}{\lambda e^{h}},\delta_R+\frac{1}{\lambda e^{h}}\right)$, but cannot guarantee that all localized states are just outside this range, as shown in Fig.~\ref{F2}(b1)(c1).

Next, let's analyze what happens when one consider the complex eigenvalue. The analytical and numerical results are in perfect agreement [see the middle column of Fig.~\ref{F2}]. The reason is that substituting $|a_2|=|E_{c}-\delta|$ for $Re(E)$ is in essence taking into account the full effect of the complex eigenvalues, not just the effect of the real part. So, one can see ME precisely divides the extended region ($|a_{2}|<\frac{1}{\lambda/e^{h}}$) and the localized region ($|a_{2}|>\frac{1}{\lambda/e^{h}}$). There are, however, one drawback to this approach, since there is a modulus operation in the process of solving ME, the ME corresponding to a negative value becomes a positive one, so chances are that information of the specific number of ME has been lost, which we will discuss at length in the section of multiple MRs. It can thus be seen that the coexistence of extended and localized states in the non-Hermitian system cannot be fully demonstrated on the real axis only, so it is necessary to extend the concept of ME to the complex plane.

As shown in the rightmost column of Fig.~\ref{F2}, MR can not only separate the extended domain from the localized domain as a boundary, but also reflect the distribution of the localized and extended state in the complex plane. To be specific, the inside of MR is the extended state, while the outside of MR the localized state. For the case of Hermitian limit ($h = 0$), MR presents itelf as a black dashed ring with the center $(1,0)$ and the radius $1/\lambda$, while MEs are red dashed lines at the intersection of MR and the real axis [see Fig.~\ref{F2}(a3)]. It is not difficult to find that under the Hermitian limit, since the eigenvalues are pure real numbers, both MR and ME can be used as boundaries to separate the extended state from the localized state in the complex plane. However, for the non-Hermitian case ($h=1$), MR, as a circle centered at $(\delta_R,\delta_I)$ with radius $1/(\lambda e^{1})$, can perfectly separates the extended and localized states inside and outside the circle, whereas MEs cannot [see Fig.~\ref{F2}(b3)(c3)].

\emph{In a word, for the non-Hermitian system, only MR can show the complete information of the coexistence of extended and localized states in the complex plane.}

\begin{figure}[tbp]
	\centering	
    \includegraphics[width=8cm]{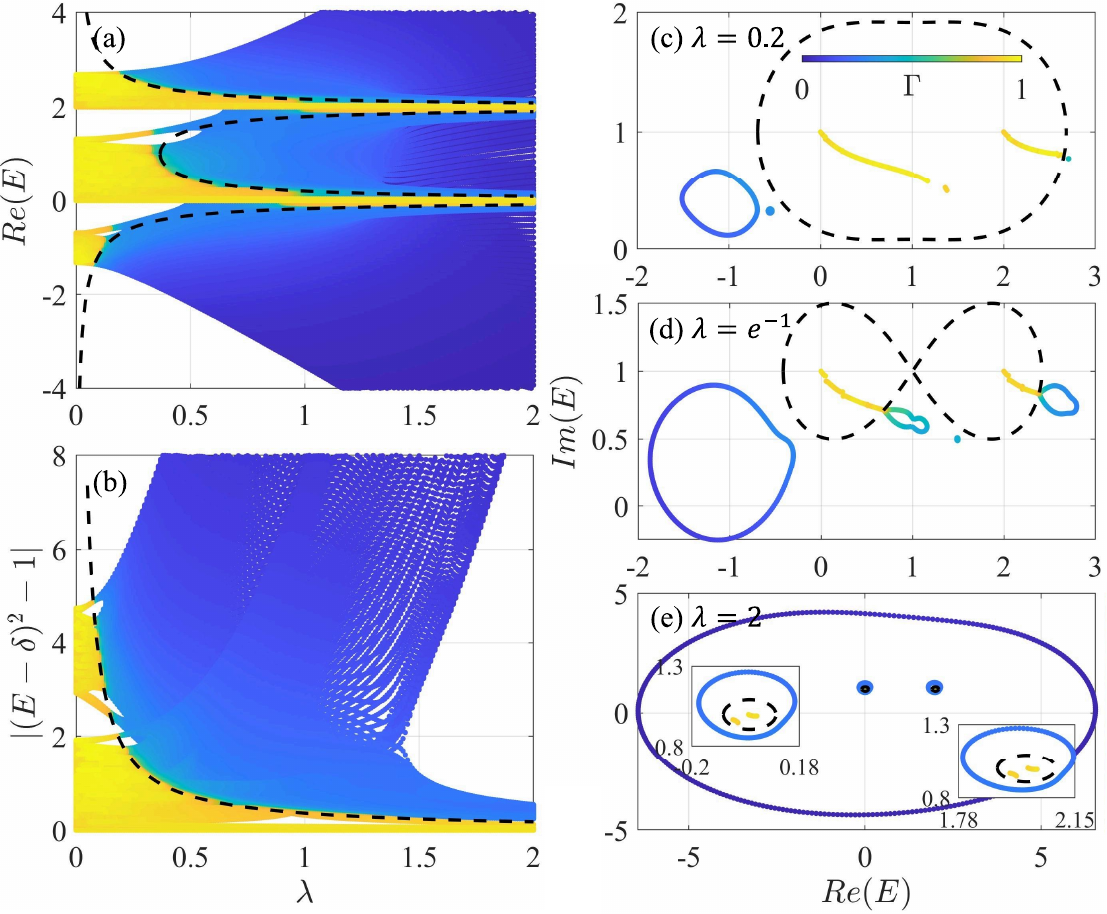}
	\caption{$\Gamma$ versus [$Re(E)$, $\lambda$] (a) and [$|a_{3}|$, $\lambda$] (b), where the dashed lines represent MEs. (c)-(e) The fractal dimension $\Gamma$ in the complex plane for $\lambda=0.2,~e^{-1},~2$. The dashed contours are MR given by Eq.~\eqref{kappa3}. Throughout, $L=987$, $\kappa=3$, $h=1$, and $\delta=1+1i$. }\label{F3}
\end{figure}

\textcolor{blue}{\emph{Multiple mobility rings.}}---When $\kappa>2$, multiple MRs will emerge in the non-Hermitian system. Fig.~\ref{F3} shows the fractal dimension $\Gamma$ corresponding to $\kappa=3$ in the case without $PT$-symmetry, where dashed lines and rings represent the MEs and MRs given by $E_{c}=\pm\sqrt{\pm\frac{1}{\lambda e^{h}}+1}+\delta_{R}$ and analytical expression~\eqref{kappa3}, respectively. Here we see again the discrepancy between the analytical and numerical MEs caused by not counting in the imaginary part. However, when we consider the complex spectrum, i.e., to calculate MR by substituting $|a_3|=|(E-\delta)^2-1|$ for $Re(E)$, it becomes a clear boundary to separate extended from localized states. Note that, information on the number of MEs will be lost because of the modulus operation during calculation, which is the price we must pay to display on one axis the influence of both real and imaginary parts on MEs [see Fig.~\ref{F3}(a)(b)].

Fig.~\ref{F3}(c)-(e) show the changes of MR with $\lambda$. One can see that MR well separates the extended and localized states, and changes in the number of MR versus $\lambda$ are also well demonstrated in the complex plane. Specifically, MR takes on an $\infty$-shape at the critical point ($\lambda_c=e^{-h}$). When $\lambda<\lambda_c$, there is only one MR in the complex plane, with the extended state distributed inside MR and the localized state outside, whereas when $\lambda>\lambda_c$, two MRs emerge in the complex plane, with the extended state distributed inside the two rings and the localized state outside. Comparing with multiple MEs in Hermitian case~\cite{YWang2020}, one can draw the following conclusions:
\emph{A non-Hermitian mosaic model will have a maximum of $\kappa-1$ MRs.}

%%%%%%
%%%%%%
%%%%%%
%%%%%%

\textcolor{blue}{\emph{The limitation of the self-duality method}}---In this part, we will clarify the limitations of duality relation in calculating the critical energy in the complex plane by comparing the critical energy obtained by Arvila's global theory and the self-duality method. We replace potential~\eqref{mosaic} by the non-Hermitian Ganeshan-Pixley-Das Sarma (GPD) potential~\cite{SGaneshan2015}
%%%
%%%
\begin{equation}\label{GAA}
V_{j}=\frac{i\lambda\cos(2\pi\alpha j+\theta)}{1-b\cos(2\pi\alpha j+\theta)},
\end{equation}
%%%
%%%
where $\lambda$ denotes the quasiperiodic intensity, $\alpha$ is the quasiperiodic parameter, $b$ is a deformation parameter, and $\theta$ denotes a global phase. We set $\alpha=(\sqrt{5}-1)/2$, $|b|<1$. It is clear that the Hamiltonian with the potential Eq.~\eqref{GAA} does not have $PT$-symmetry and when $\lambda>0$ the system will have a complex spectrum. The corresponding Hamiltonian can be obtained by successive transformations with its exact self-duality relation $E_{c}=(\pm 2-i\lambda)/b$~\cite{Sup}. One can obtain the self-duality points by the hidden duality method~\cite{Sup,MGoncalves2022,MGoncalves2023a,MGoncalves2023b}. The irrational number $\alpha$ can be approximated by the Fibonacci number $F_{m}/F_{m+1}$, i.e., $\alpha =(\sqrt{5}-1)/2$ for $m\rightarrow\infty$. For the lowest order approximation, $\alpha = 1$, the eigenequation for the Hamiltonian reads
%%%
%%%
\begin{equation}\label{Eigeq}
J(\psi_{j-1}+\psi_{j+1})+\frac{i\lambda\cos(\theta)}{1-b\cos(\theta)}\psi_{j}=E\psi_{j}.
\end{equation}
%%%
%%%
By the Fourier transform, one can get its eigenvalue as 
%%%
%%%
\begin{equation}
\begin{aligned}
&E(\theta,k)  = \frac{i\lambda\cos(\theta)}{1-b\cos(\theta)}+2\cos(k)\\
&\Rightarrow E=(bE+i\lambda)\cos(\theta)+2\cos(k)-2b\cos(k)\cos(\theta)
\end{aligned}
\end{equation}
%%%
%%%
for $1-b\cos(\theta)\neq 0$, which is self-duality under $\theta\leftrightarrow k$ if $E_{c,1} = (2 - i\lambda)/b$. Defining $\theta'= \theta + \pi$, we get $E_{c,2} = -(2 + i\lambda)/b$. This result is the same as the exact self-duality relation. In contrast, the LE obtained by Avila's global theory is~\cite{Sup}
%%%
%%%
\begin{equation}\label{LEEGAA}
\gamma(E)=\mathrm{max}\left\{\ln\left|\frac{Eb+i\lambda\pm\sqrt{(Eb+i\lambda)^2-4b^2}}{2(1+\sqrt{1-b^2})}\right|,0 \right\}.
\end{equation}
%%%
%%%
Let $E=E_{R}+iE_{I}$, we can obtain the exact expression of MR
%%%
%%%
\begin{equation}\label{MRGAA}
\frac{(bE_{R})^2}{4}+\frac{(bE_{I}+\lambda)^2}{C^2}=1
\end{equation}
%%%
%%%
for $\gamma(E)=1$, where $C=[(1+\sqrt{1-b^2})^2-b^2]/(1+\sqrt{1-b^2})$.
%%%%%%
%%%%%%
 \begin{figure}[tbp]
	\centering	
    \includegraphics[width=7cm]{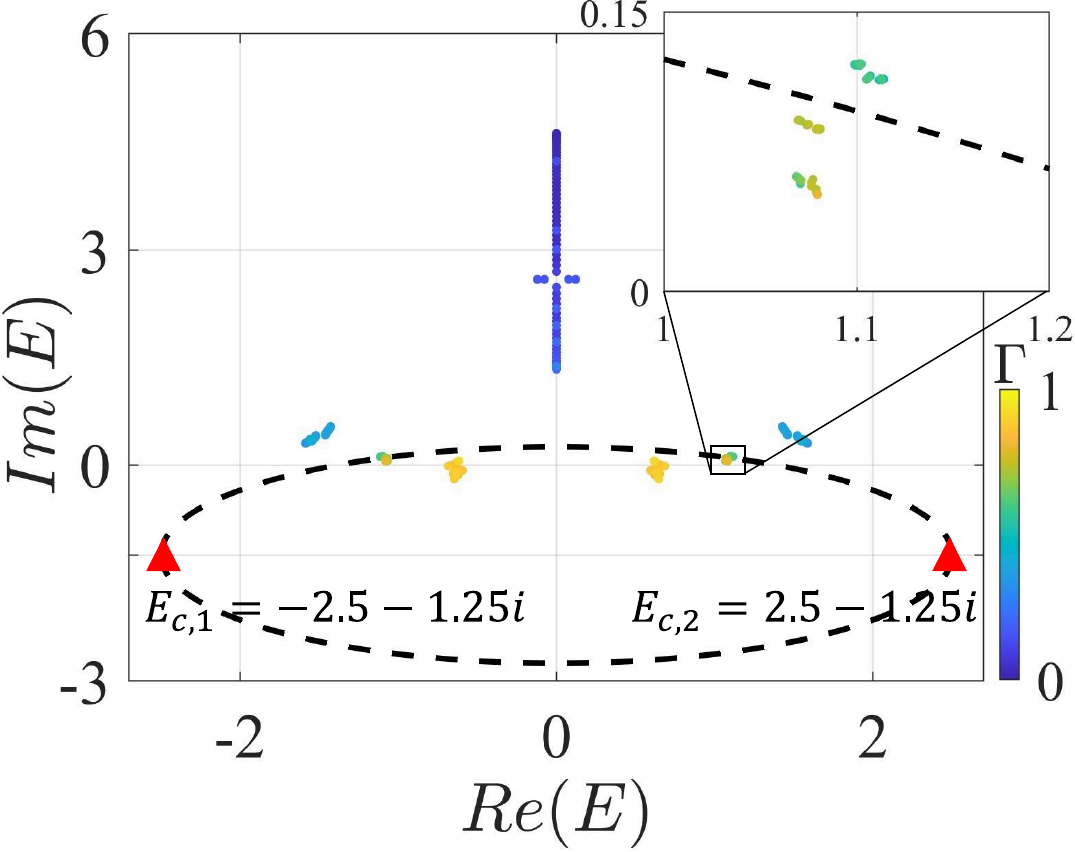}
	\caption{$\Gamma$ in the complex plane. The dashed circle denotes the LE given by Eq.~\eqref{LEEGAA}, whereas the two red triangles arecself-duality points obtained from the self-duality relation. Throughout, $L=610$, $\lambda=1$, $b=0.8$ and $\theta=0$. }\label{b05}
\end{figure}
%%%%%%
%%%%%%

Fig.~\ref{b05} shows the fractal dimensions in the complex plane, where the black dashed ring is MR obtained by Avila's global theory (the elliptic equation), and the red triangles are the energy critical points obtained by the self-duality relation. MR perfectly divides the extended and the localized regions, and shows how the extended and the localized states distribute in the complex plane, i.e., the extended (localized) states are inside (outside) MR. However, the self-duality relation can merely obtain two special critical points on MR, namely, $Im(E)=\lambda/b$. This result follows from the existence of an infinite number of self-dual points in non-Hermitian systems. \emph{That is to say, the self-duality relation is not comprehensive in studying the MEs in the non-Hermitian system.}

\textcolor{blue}{\emph{Conclusion.}}---In conclusion, MEs in the non-Hermitian system are studied. Since they exhibit a ring structure in the complex plane, we call them ``mobility ring''. Further, we analyze MR in various cases and find that MR is universal for the non-Hermitian system, no matter the system has $PT$-symmetry or not. In addition, by analogy with multiple MEs in the Hermitian system, we investigate multiple MRs in complex plane, and give the expression of the maximum number of MRs in the mosaic model. Finally, by comparing the results given by Arvila's global theory and self-duality relation, we find that self-duality relation has limitations in studying the critical energy in the complex plane. Generally, non-Hermitian systems have complex spectra, which means that MR studied in this paper is universal in the complex plane. Hopefully our findings will bring benefits to the study of ME theory and non-Hermitian physics.

\textcolor{blue}{Note added:} In completing this manuscript, we note that a recent preprint entitled ``Non-Hermitian butterfly spectra in a family of quasiperiodic lattices'', similarly investigates the mobility edges in non-Hermitian system~\cite{LWang2024}. Their article is from the perspective of butterfly spectrum. In this paper, in addition to studying the spectrum, we further discuss the properties of multiple mobility rings and clarify the limitations of self-duality method in non-Hermitian cases.

\textcolor{blue}{\emph{Acknowledgements.}}---We thank Qi Zhou for his insightful suggestions. This work was supported by the National Key Research and Development Program of China (Grant No.2022YFA1405300), the National Natural Science Foundation of China (Grant No.12074180), and the Guangdong Basic and Applied Basic Research Foundation (Grants No.2021A1515012350).

%\bibliographystyle{apsrev4-1}
%\bibliography{cite.bib}
\global\long\def\id{\mathbbm{1}}
\global\long\def\ui{\mathbbm{i}}
\global\long\def\ud{\mathrm{d}}

%%%%%%%%%%%%%%%%%%%%%%%%%%%%%%%%%%%%%%%%%%%%%%%%%%%%%%%%%%%%%%%
\setcounter{equation}{0} \setcounter{figure}{0}
\setcounter{table}{0} %\setcounter{page}{1} \makeatletter
\renewcommand{\theparagraph}{\bf}
\renewcommand{\thefigure}{S\arabic{figure}}
\renewcommand{\theequation}{S\arabic{equation}}

\onecolumngrid
\flushbottom
%%%%%%%%%%%%%%%%%%%%%%%%%%%%%%%%%%%%%%%%%%%%%%%%
\newpage
\section*{Supplementary materials for:\\ ``Ring Structure in the Complex Plane: A Fingerprint of non-Hermitian Mobility Edge''}
\section{I. Lyapunov exponents in the complex plane}\label{SSII}
\begin{figure}[htbp]
	\centering	
    \includegraphics[width=10cm]{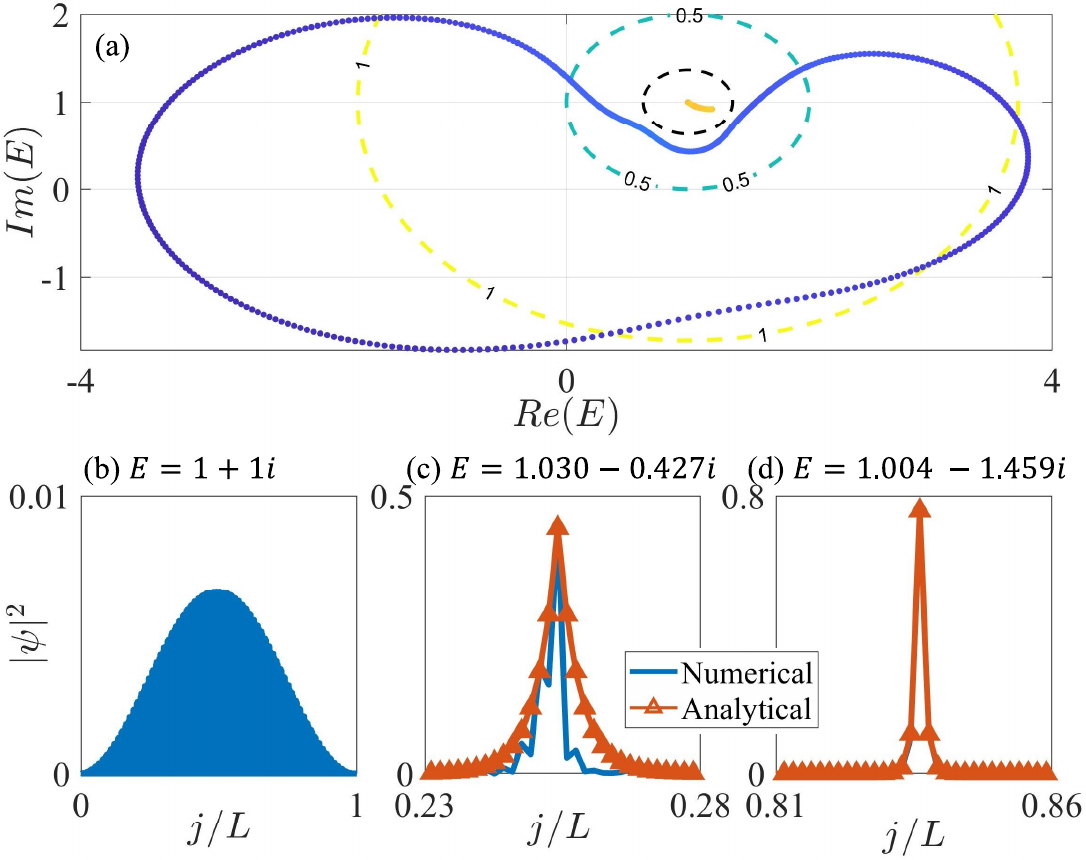}
	\caption{The fractal dimension $\Gamma$ of the eigenstates corresponding to the eigenvalues $E$ in the complex plane for $\kappa=2$, where the black rings are MR and the contours of LE are given by Eq.~\eqref{gamma}. The eigenvalues are the distributions of eigenstates over all lattices corresponding to (a) $E =1+1i$, (b) $E=1.030-0.427i$, and (c) $E =1.004-1.459$, where the orange triangle is given by the fitted Eq.~\eqref{psi}. We fix $L=610$, $\delta=1+i$, $h=1$, and $\lambda=1$. }
 \label{FS1}
\end{figure}

If the energy $E$ corresponds to an eigenstate that is localized, then its localized length $\Lambda=1/\gamma$ and the wave function of the eigenstate $\psi$ can be given by
%%%
%%%
\begin{equation}\label{psi}
|\psi|\propto |\psi|_{\max}e^{-\gamma|j-j_{c}|},
\end{equation}
%%%
%%%
where $|\psi|_{\max}$ is the maximum value of $|\psi|$ at the center site $j_c$ of localization. In Fig.~\ref{FS1}(a) we show the fractal dimensions $\Gamma$ of all eigenstates in the complex plane for $\delta = 1 + i$, $h = 1$, and plot the contours of the LE via $2\gamma(E)=\max\left\{\ln|\lambda a_{2}|+|h|,0\right\}$. The extension under non-Hermitian conditions originates from LE in the complex plane. The region with $\gamma= 0$ forms one valley in the complex plane (inside the MR), and the eigenstate $\Gamma\rightarrow1$ in this region is the extended state. The further away the eigenstate is from MR, the smaller $\Gamma$ is, and the more localized the corresponding eigenstate is. For eigenstates within MR, as shown in Fig.~\ref{FS1}(b), the wavefunction can extend to all lattices. For localized states with $\gamma > 0$, as shown in Fig.~\ref{FS1}(c)(d), we present the distributions of the eigenstates for $\gamma=0.222$ and $\gamma=0.950$ over all lattices, respectively, and compare them with the fitted Eq.~\eqref{psi}. It is evident that Eq.~\eqref{psi} effectively characterizes the localization features of the eigenstates, and the larger LE becomes, the more localized it is.

%%%%%%
%%%%%%
%%%%%%
%%%%%%

\section{II. Non-Hermitian Ganeshan-Pixley-Das Sarma model}\label{SSI}
\subsection{A. Exact self-duality relation}
Here, we show that the mobility edge (ME) obtained from the self-duality relation does not give a good indication of the MEs under the non-Hermitian system. We consider a non-Hermitian GPD model:
%%%
%%%
\begin{equation}\label{S1}
H=\sum_{j=1}^{L-1}J(c_{j}^{\dagger}c_{j+1}+\mathrm{H.c.})+\sum_{j=1}^{L}V_{j}c_{j}^{\dagger}c_{j}, 
\end{equation}
%%%
%%%
where
%%%
%%%
\begin{equation}\label{S2}
V_{j}=\frac{i\lambda\cos(2\pi\alpha j+\theta)}{1-b\cos(2\pi\alpha j+\theta)}, 
\end{equation}
%%%
%%%
$c_{j}^{\dagger}$ ($c_{j}$) creates (annihilates) a fermion on site $j$, and $\mathrm{H.c.}$ stands for the Hermitian conjugate. $J$ represents the nearest neighbor hopping strength. The quasiperiodic potential is regulated by the following parameters: $\lambda$ denotes the quasiperiodic intensity, $\alpha$ is the quasiperiodic parameter, $b$ is a deformation parameter, and $\theta$ denotes a global phase. We set $\alpha=(\sqrt{5}-1)/2$, $|b|<1$, and $J=1$. It is clear that the Hamiltonian~\eqref{S1} does not have $PT$-symmetry and the system will have a complex spectrum. Next, we follow the generalized self-duality transformation to obtain the MEs under duality relations.

The eigenequation for the Hamiltonian quantity~\eqref{S1} can be written as
%%%
%%%
\begin{equation}\label{S3}
J(\psi_{j-1}+\psi_{j+1})+V_{j}\psi_{j}=E\psi_{j}.
\end{equation}
%%%
%%%
We define
%%%
%%%
\begin{equation}\label{S4}
\chi_{j}(\omega,\theta)=\frac{\sinh{\omega}}{\cosh{\omega}-\cos(2\pi\alpha j+\theta)},
\end{equation}
%%%
%%%
where $\omega=1/b$. Thus, the Eq.~\eqref{S3} can be written directly as
%%%
%%%
\begin{equation}\label{S5}
J(\psi_{j-1}+\psi_{j+1})+G\chi_{j}(\omega,\theta)\psi_{j}=(E+i\lambda\cosh{\omega})\psi_{j},
\end{equation}
%%%
%%%
in which $G=i\lambda\cosh\omega\coth\omega$. By using a well-established mathematical relation as following,
%%%
%%%
\begin{equation}
\frac{\sinh{\omega}}{\cosh{\omega}-\cos(2\pi\alpha j+\theta)}=\sum_{r=-\infty}^{\infty}e^{-\omega|r|}e^{ir(2\pi\alpha+\theta)}.
\end{equation}
%%%
%%%
Define $u_{p}=\sum_{j=-\infty}^{\infty}e^{ij(2\pi\alpha p+q\pi)}$ and $q$ is an integer. Multiplying $e^{ij(2\pi\alpha p+q\pi)}$ with both sides of Eq.~\eqref{S5} and performing a summation, we get
%%%
%%%
\begin{equation}\label{S7}
\eta \chi_{p}^{-1}(\omega_0,0)e^{p\theta}u_{p}=G\sum_{r}e^{-\omega|p-r|}e^{r\theta}u_{r},
\end{equation}
%%%
%%%
where $\omega_0$ is defined through relation $E+i\lambda\cosh{\omega}=(-1)^{q}2J\cosh{\omega_0}$ and $\eta=(-1)^{q}2J\sinh{\omega_0}$. By multiplying $v_{m}=\sum_{p}e^{ip(2\pi\alpha m+\theta q\pi)}\chi_{p}^{1}(\omega_0,0)u_{p}$ with both sides and making a sum over $p$, Eq.~\eqref{S7} is correspondingly transformed into
%%%
%%%
\begin{equation}\label{S8}
\eta \chi_{m}^{-1}(\omega,q\pi)v_{m}=G\sum_{r}e^{-\omega_{0}|m-r|}v_{r}.
\end{equation}
%%%
%%%
In the last step we define $z_{k}=\sum_{m}e^{im(2\pi\alpha k+\theta)}v_{m}$. We multiply Eq.~\eqref{S8} by $e^{im(2\pi\alpha j+\theta)}$ and sum over $m$ to obtain 
%%%
%%%
\begin{equation}\label{S9}
J(z_{k-1}+z_{k+1})+G\frac{\sinh{\omega}}{\sinh{\omega_0}}\chi_{k}(\omega_0,\theta)z_{k}=(-1)^{q}2J\cosh{\omega}z_{k}.
\end{equation}
%%%
%%%
Comparing Eq.~\eqref{S9} and Eq.~\eqref{S5}, if we make $\omega = \omega_0$, the two equations are completely equivalent. Thus, we have $E+i\lambda\cosh{\beta}=(-1)^{q}2J$, which in terms of the original parameter $b$ is
%%%
%%%
\begin{equation}\label{S10}
bE=(-1)^{q}2J-i\lambda.
\end{equation}
%%%
%%%
Since $q$ is an integer, it has a pair of MEs $E_{c}=(\pm 2J-i\lambda)/b$. It can be seen that the self-duality relation is such that only two critical points in the complex plane can be given. 

\subsection{B. Hidden duality}
This self-duality point is also able to be obtained by hidden duality. First, we consider a generic potential $V(2\pi\alpha j)$ with a period $V(x+2\pi)=V(x)$ of $2\pi$. If $\alpha$ is irrational, then $V(x)$ is exact but never repeated. However, for irrational numbers we can approximate $\alpha=n_{1}/n_{2}$ by two rational numbers. For the golden ratio $\alpha=(\sqrt{5}-1)/2$, we can approximate it by the Fibonacci numbers $\alpha=F_{m-1}/F_{m}$ and only if $m\rightarrow\infty$, $\alpha = (\sqrt{5}-1)/2$. When $m$ is small, we can obtain the self-duality relation by Bloch's theorem. The approximation will become more and more accurate as $m$ keeps increasing. For arbitrary quasiperiodic models, duality can also be established by means of the commensurate approximation. Assuming that $\alpha=n_1/n_2$, the system will give an expansion to $n_2$ sites as a unit cell and can be solved in terms of rescaled Bloch momentum $K=k/n_{2}\in [-\pi,\pi]$. Because of the translational invariance under shifting $n_2$, the phase $\theta$ is also rescaled into $\phi=\theta/n_{2}\in [-\pi,\pi]$. We can then solve the Fermi surface in the two-dimensional $(K,\phi)$ phase by solving the matrix
%%%
%%%
\begin{equation}
H(\phi,K)=\begin{pmatrix}
V(\phi/n_{2})  & 1 & \dots & e^{iK}  \\
1   & V(2\phi n_{1}/n_{2}+\phi/n_{2}) &   \dots &0  \\
 \vdots &  & \ddots  &  \vdots \\
e^{-iK}  &  \dots &1  & V(2\phi n_{1}(n_{2}-1)/n_{2})+\phi/n_{2})  \\
\end{pmatrix}.
\end{equation}
%%%
%%%
On the self-duality point, its Fermi surface at $n_{1},n_{2}\rightarrow\infty$ is invariant under $\phi\leftrightarrow K$ and $\phi'\leftrightarrow K$, where $\phi'=\phi+\pi$. For some models, they have duality properties for any $n_2$. We start from the simplest $n_2 = 1$, i.e., $\alpha=1$. For the AAH model $V_{j}=2\lambda\cos(2\pi\alpha j+\theta)$, the eigenvalues are
%%%
%%%
\begin{equation}
H(\phi,K)=2\lambda\cos(\phi)+2\cos(K).
\end{equation}
%%%
%%%
It can be clearly seen that under the transformation of $\phi\leftrightarrow K$ ($\phi'\leftrightarrow K$), the eigenvalue remains unchanged when $\lambda=1$ ($\lambda=-1$). Putting the two conditions together, we get the self-duality condition $|\lambda| =1$. 

For the case with MEs, the dispersion relation for the eigenvalues of Hermitian's GPD model $V_{j}=\lambda\cos(2\pi\alpha j+\theta)/(1-b\cos(2\pi\alpha j+\theta))$, as an example, is
%%%
%%%
\begin{equation}
E(\phi,K)  = \frac{\lambda\cos(\phi)}{1-b\cos(\phi)}+2\cos(K)\Rightarrow E=(bE+\lambda)\cos(\phi)+2\cos(K)-2b\cos(K)\cos(\phi).
\end{equation}
%%%
%%%
By $\phi\leftrightarrow K$ and $\phi'\leftrightarrow K$, we can obtain its self-duality point as $E=(\pm 2-\lambda)/b$ and this result is in agreement with Ref.~\cite{SGaneshan2015}. In addition, this method works well in solving models that do not have self-duality.

\subsection{C. Avila's global theory}
Further, we contrast this by Avila's global theory. we can calculate LE from the transfer matrix
%%%
%%%
\begin{equation}\label{LE}
\gamma_{\epsilon }(E)=\lim_{L\rightarrow \infty }\frac{1}{2\pi N}\int  \ln \left \| T_{L}(\theta )  \right \|d\theta ,
\end{equation}
%%%
%%%
where $T_{L}=\prod_{j=1}^{L} T_{n}=T_{L}T_{L-1}\cdots T_{2}T_{1}$ and $\left \| \cdot  \right \|$ denotes a matrix norm. In the thermodynamic limit $L\rightarrow\infty$, the extended state has $\gamma = 0$, while the localized state has $\gamma > 0$. From the eigenequation~\eqref{S3} we can write the transfer matrix
%%%
%%%
\begin{equation}
T_{j}=\begin{pmatrix}
E-\frac{i\lambda\cos(2\pi\alpha j+\theta)}{1-b\cos(2\pi\alpha j+\theta)}  &-1 \\
 1 &0
\end{pmatrix}=A_{j}B_{j},
\end{equation}
%%%
%%%
where 
%%%
%%%
\begin{equation}
A_{j}=\frac{1}{1-b\cos(2\pi\alpha j+\theta)}
\end{equation}
%%%
%%%
and
%%%
%%%
\begin{equation}
B_{j}=
\begin{pmatrix}
E[1-b\cos(2\pi\alpha j+\theta)]-i\lambda\cos(2\pi\alpha j+\theta)  &-1+b\cos(2\pi\alpha j+\theta) \\
1-b\cos(2\pi\alpha j+\theta) &0
\end{pmatrix},
\end{equation}
%%%
%%%
Then, $\gamma(E)=\gamma_{A}(E)+\gamma_{B}(E)$, where $\gamma_{A}(E)=\lim_{j\rightarrow \infty}\frac{1}{2\pi L}\int \ln\left \|A_{L}(\theta)  \right \|d\theta$. Next, we extend the phase $\theta\rightarrow\theta+i\epsilon$. Due to the ergodicity of the map $\theta\rightarrow 2\pi\alpha j+\theta$, we can write $\gamma_A(E)$ as integral over phase $\theta$, which results in
%%%
%%%
\begin{equation}\label{S15}
\gamma_{A}(E)=-\frac{1}{2\pi}\int_{0}^{2\pi}\ln\left|\frac{1}{1-b\cos(2\pi\alpha j+\theta+i\epsilon)}\right|d\theta=-\ln\left(\frac{1+\sqrt{1-b^2}}{2}\right)
\end{equation}
%%%
%%%
when $|b|<1$ and $\epsilon<\ln\left(\frac{1+\sqrt{1-b^2}}{2}\right)$. For $\gamma_{B}(E)$, we take $\epsilon\rightarrow\infty$ to get
%%%
%%%
\begin{equation}
B_{j}(E,\epsilon)=\frac{e^{-(2\pi\alpha j+\theta)+\epsilon}}{2}\begin{pmatrix}
 -(bE+i\lambda) & b\\
 -b &0
\end{pmatrix}+O(1).
\end{equation}
%%%
%%%
According to the Avila's global theory, $\gamma_{B}(E)$ is a convex, piecewise linear function with integer slopes. Thus, the LE about $B_{j}(E,\epsilon)$ can be written as
%%%
%%%
\begin{equation}
\gamma_{B}(E,\epsilon)=\epsilon+\left|\frac{Eb+ i\lambda\pm\sqrt{(Eb+i\lambda)^2-4b^2}}{4}\right|
\end{equation}
%%%
%%%
for any $\epsilon\in \left(-\infty,\infty\right)$. Based on Eq.~\eqref{S15} and the non-negativity of LE, we have
%%%
%%%
\begin{equation}\label{S18}
\gamma(E,0)=\mathrm{max}\left\{\ln\left|\frac{Eb+ i\lambda\pm\sqrt{(Eb+i\lambda)^2-4b^2}}{2(1+\sqrt{1-b^2})}\right|,0 \right\}.
\end{equation}
%%%
%%%
Then ME can be determined by $\gamma(E)=0$. Let $E=E_{R}+iE_{I}$, we can obtain the MR
%%%
%%%
\begin{equation}\label{MRGAA}
\frac{(bE_{R})^2}{4}+\frac{(bE_{I}+\lambda)^2}{C^2}=1
\end{equation}
%%%
%%%
for $\gamma(E)=1$, where $C=[(1+\sqrt{1-b^2})^2-b^2]/(1+\sqrt{1-b^2})$. It is obviously an ellipse.

\section{III. Localized phase transitions for non-Hermitian GPD models}
Here, we show the fractal dimension $\Gamma$ of the non-Hermitian GPD model for the remaining parameters. As can be seen in Fig.~\ref{S2}(a)-(c) for $b = 0.4$ and (d)-(f) for $b = 0.8$, MR divides well the extended and localized states in the complex plane.

%%%%%%
%%%%%%
 \begin{figure}[htbp]
	\centering	
    \includegraphics[width=16cm]{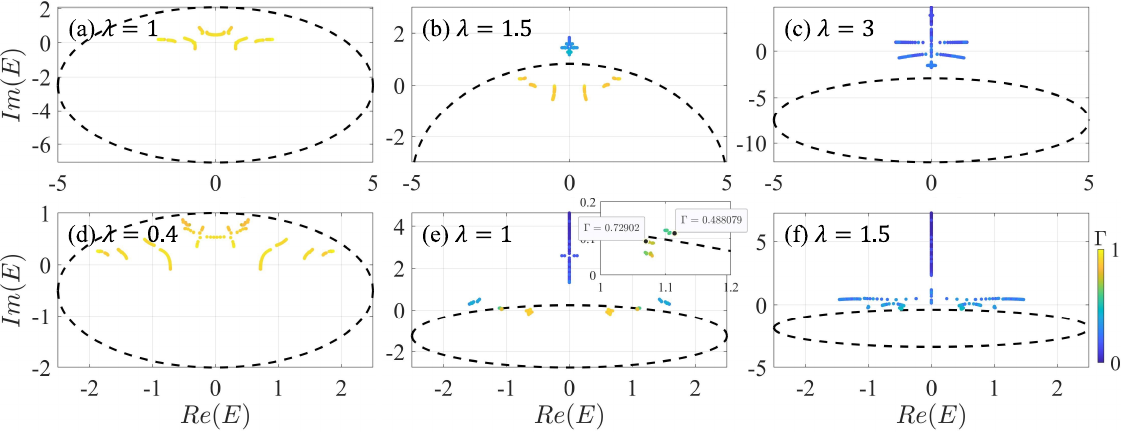}
	\caption{(a)-(c) $b=0.4$ and (d)-(e) $b=0.8$ for the fractal dimension $\Gamma$ of all eigenstates in the complex plane under different $\lambda$, where black ring is the MR. We fix $L=610$ and $\theta=0$.}\label{S2}
\end{figure}
%%%%%%
%%%%%%

\end{document}